\documentclass[a4paper,usenatbib,fleqn]{mnras}

\makeatletter

\newcommand{\Rmnum}[1]{\expandafter\@slowromancap\romannumeral #1@}
\makeatother

\usepackage{graphicx}   
\usepackage{amsmath}    
\usepackage{amssymb}    
\usepackage{subcaption}
\usepackage{color,soul}
\usepackage{xcolor}
\usepackage{newtxtext,newtxmath}

\usepackage{color}

\def\prd{{\rm Phys. Rev. D}}
\def\apj{{\rm ApJ}}
\def\aj{{\rm AJ}}

\def\mnras{{\rm MNRAS}}

\def\hmpc{h^{-1}{\rm Mpc}}

\def\kms{\, {\rm km}\, {\rm s}^{-1}}

\def\lya{Ly$\alpha$ }

\def\simlt{\lower.5ex\hbox{$\; \buildrel < \over \sim \;$}}
\def\simgt{\lower.5ex\hbox{$\; \buildrel > \over \sim \;$}}

\title[Deep Forest]{
  Deep Forest: Neural Network reconstruction of the Lyman-$\alpha$ forest.
}

\author[Huang et al.]{\parbox{18cm}{Lawrence Huang$^{1,2}$,
Rupert A.C. Croft$^{1,2}$\thanks{E-mail: rcroft@cmu.edu},
and Hitesh Arora$^{3}$
}\vspace{0.3cm}\\
$^{1}$ McWilliams Center for Cosmology, Dept. of Physics, 
Carnegie   Mellon  University, Pittsburgh, PA 15213, USA\\
$^{2}$ NSF AI Planning Institute for Physics of the Future, 
Carnegie   Mellon  University, Pittsburgh, PA 15213, USA\\
$^{3}$ Robotics Institute, School of Computer Science,
Carnegie Mellon University,
Pittsburgh, PA 15213, USA
}

\begin{document}

\topmargin=-1.0cm

\maketitle


\begin{abstract}

We explore the use of Deep Learning to infer physical quantities from
the observable transmitted flux in the \lya forest. We train a
Neural Network using redshift $z=3$ outputs from cosmological hydrodynamic simulations and
mock datasets constructed from them. We
evaluate how well the trained network is able to reconstruct the optical
depth for \lya\ forest absorption from noisy and often saturated transmitted
flux data. The Neural Network outperforms an alternative reconstruction method
involving log inversion and spline interpolation by approximately a factor of 2 in the
optical depth root mean square error. We find no significant 
dependence in the improvement on input data signal to noise, although the gain is
greatest in high optical depth regions. The \lya forest optical depth studied here serves as
a simple, one dimensional, example but the use of Deep Learning and simulations to approach
the inverse problem in cosmology could be extended to other physical quantities and
higher dimensional data.

\end{abstract}


\section{Introduction}
\label{intro}

The $\Lambda$CDM cosmology \citep[e.g.,][]{dodelson03}, combined with numerical
simulations (see the review by \citealt{vogelsberger20}) can be used to create realistic and detailed forward
models. Some observables such as the Lyman-$\alpha$ forest (\citealt{rauch98}, \citealt{weinberg03}) are
particularly useful because almost all the relevant physical processes
are understood and can be resolved (\citealt{cen94,zhang95,hernquist96,hui97}). Given this level of fidelity, an
interesting question is how these forward models can be used in conjunction
with
observational data in order to infer unobservable quantities, such as
the dark matter distribution from galaxy positions, cool gas using observations
of hot gas, or even the initial density fluctuations from data at redshift
zero. The advent of efficient  machine learning algorithms (see e.g., \citealt{mitchell97})
offers a route to solving this inverse problem, and one that we explore in this paper.
In particular, we will use Deep Learning (DL, \citealt{lecun15}), the
science of neural networks (NN), combined with numerical
simulations. We will train NN using simulations that have well defined inputs
and outputs. We will then use those networks to infer the output (underlying
physical quantity) given an input (observational data).

The use of DL techniques in cosmology and astrophysics has exploded
over the last few years, following the trend of increasing application
of Artificial Intelligence (AI) to many scientific fields and to
everyday life \citep{russell2020}. With DL, artificial Neural Networks are
used that are capable of learning, including from data that is unstructured or unlabeled. The NNs consist of 
neurons arranged in layers, with numerical  values passed between neurons
subjected to weights which are adjusted as part of the training process.
An introduction to DL and NN is \cite{goodfellow16}. 
Their use in astronomy so far has often been  to find
and classify events, and example training in this case consists in
providing labelled  datasets, with the NN learning to associate
particular inputs (for example astronomical images) with output labels
(e.g., galaxy types, \citealt{cheng20}). Lyman-$\alpha$ forest data is one dimensional,
 and DL has been used successfully in one dimension to find Gravitational
 Wave events from strain time series \citep{george2018}, to classify astronomical
 spectra \citep{muthukrishna19}, and also to find and characterize high column density
 absorption lines in quasar spectra
 \citep{parks18}. Applications to two
 dimensional images are more common (e.g., finding gravitational
 lenses \citep{metcalf19}, or adding subresolution details to galaxy
 images \citep{schawinski17}).
 
 More recently, DL techniques are being applied increasingly widely to the simulation of datasets and to the analysis of data. NN trained using a grid of N-body simulations have been used to infer cosmological parameters from galaxy weak lensing maps by \cite{fluri2019}. Maps of the lensing potential itself have been reconstructed from CMB observations (\citealt{caldeira2019}), an example of a DL solution to an inverse problem similar to the type we consider here. Also closely related
 is the work of \cite{charnock2018} who use Information Maximizing NN to optimally compress
 data, and show as an example cosmological constraints inferred from quasar spectra (specifically the
 Lyman-$\alpha$ forest, see below). Training sets derived from simulations feature heavily in this
 work, but DL has also been used to interpret and learn the physical processes occurring in the 
 simulations (e.g., \citealt{he2019}, \citealt{lucie2018}), as well as becoming part of the simulation methodologies themselves (e.g., \citealt{ramanah20}, \citealt{li20}).

The
physical system we will concentrate on is the Lyman-$\alpha$ 
forest of absorption due to neutral hydrogen seen in quasar
and galaxy spectra \citep{rauch98,savaglio02}, because the physics is
well understood (as mentioned above) and also because
the observations are one dimensional,
and therefore numerically easy to process.
At redshifts where the Ly$\alpha$ transition is at optical wavelengths, the
forest absorption mostly arises in the moderately overdense
intergalactic medium (IGM) \citep{bi93,cen94,zhang95,hernquist96}.
In the standard cosmological model, the forest is generated by residual
neutral hydrogen in this photoionized medium.
The space between galaxies is filled with this
absorbing material, and its structure on
scales larger than the Jean’s scale traces the overall matter density.
The relevant physics was first described by \cite{gunn65} in the context of a uniform medium, leading to the characterization in the forest as the ‘fluctuating Gunn–Peterson effect’ (FGPA, \citealt{weinberg98}). The \lya forest has been used to test cosmological models, allowing for example the measurement of the baryonic oscillation scale at redshifts $z>2$ (e.g., \citealt{aubourg15}).

The matter density, temperature, and velocity field in simulations can
be used to predict Ly$\alpha$ forest observables as mentioned above. The inverse procedure, reconstruction of
these underlying physical quantities from observations can also be carried out
\citep{nusser99,horowitz19,muller20}, although non-linearities
and incomplete information make this difficult. While the methods in
this paper could be used to carry out such reconstruction, we will
instead restrict
ourselves to a more limited problem in this first use case. We will infer
the optical depth for absorption $\tau$ by neutral hydrogen from the
transmitted flux $F$ observed in a spectrum. These quantities are
related by
\begin{equation}
\label{Feq}
F=e^{-\tau}
\end{equation}
The flux is often
saturated (particularly at high redshift), meaning that $\tau$ cannot be
directly inferred from observations of $F$.

We note that in truly dense regions, close to and in galaxies, the FGPA
is not obeyed. These are known as Lyman limit and Damped Ly$\alpha$
(DLA) systems (see e.g., \citealt{wolfe05} for a review),
 because of absorption of light
 beyond the Lyman limit and presence of damping wings respectively. These systems are however
 rare, and we will not deal with them here. Our work could
 be adapted to deal with them too, given simulations that model them
(e.g., \citealt{pontzen08}). Previous
 work has used Machine Learning techniques to detect and
 characterize them in observational data (e.g., \citealt{parks18}),
 as well as simulating them with generative NN \citep{zamudio19}.

Here we will use cosmological simulations which
resolve the relevant physics for the \lya\ forest  to make training
spectra. Once trained, NN will 
recover the optical depth $\tau$ from
the observed transmitted flux $F$.
The NN  will therefore be using information
from observable regions to infer the situation in
unobservable (saturated) regions.
We will test the fidelity
of this recovery using simulations for which both quantities
are available. Tests with different noise levels will be important as these
will dictate the fraction of spectra that are effectively saturated.
We will compare this DL recovery of optical depths to an alternative which
is to smooth spectra until they are more easily invertible directly (using Equation \ref{Feq}), along with spline interpolation for regions that are still saturated.
We  concentrate on relatively poor input signal to noise ratio (S/N) of 2.5-10 per pixel
as these are most relevant for large surveys (e.g., \citealt{lopez16}, \citealt{eboss20}).

Our plan for the paper is as follows. In Section \ref{sim} we introduce the cosmological hydrodynamic simulation \lya forest
data we use for training and testing. In Section \ref{method} we describe the NN based method we will use for reconstruction,
including data preprocessing and the network architecture. We also give details of some alternative reconstruction methods
we will use for comparison. In Section \ref{results} we present the results of our reconstructions, with both  example 
sightlines shown as well as statistical measurements of accuracy. In Section \ref{summary} we summarize our work and discuss
the results and possible future directions.

\section{The Lyman-$\alpha$ forest: training data}
\label{sim}

\subsection{Hydrodynamic
simulation}

In order to make training data for our NN, we use the \lya\ spectra computed from a large hydrodynamic
cosmological simulation of the $\Lambda$CDM model. The smoothed
particle hydrodynamics code {\small P--GADGET} (see
\citealt{springel05}, \citealt{dimatteo12}) evolved
 $2 \times 4096^2 = 137$ billion particles in a cubical periodic 
volume of $(400 \hmpc)^3$. This simulation was previously
used in other works such as \cite{cisewski14} and \cite{croft18}, where more details
are given.

 The 
cosmological parameters used
in the simulation were 
$h = 0.702,\, \Omega_{\Lambda} = 0.7
25,\, \Omega_m = 0.275, \, \Omega_b = 0.046, \, n_s = 0.968\; $and$ \;
\sigma_8 = 0.82.$ The mass per particle was $1.19 \times 10^7$ $h^{-1}
M_{\odot}$ (gas) and $5.92 \times 10^7$ $h^{-1} M_{\odot}$ (dark
matter).  An ultraviolet background radiation field consistent with that of
\cite{haardt96} is included, as well
as cooling and star formation. The star formation
model however
 uses a lower density threshold ($\rho=1000$, in units
 of the mean density) than usual
(for example in \citealt{springel03}) so that gas
particles are quickly converted to collisionless gas particles.  In this way, the execution of the simulation
is sped up, but this has no significant effect on the diffuse IGM that
gives rise to the Lyman-$\alpha$ forest.

\subsection{Mock observational data}
\label{mockobs}

 We use the simulation snapshot at redshift $z = 3.0$  
to generate a set of Lyman-$\alpha$ spectra using information
from the particle distribution \citep{hernquist96}.
The spectra are generated on a grid with $256^2 = 65536$ 
evenly spaced sightlines. These many sightlines are
therefore available for training purposes. Because neighbouring sightlines arise in the same large scale structures, they are not completely independent datasets.

\begin{figure}
    \includegraphics[width=\linewidth]{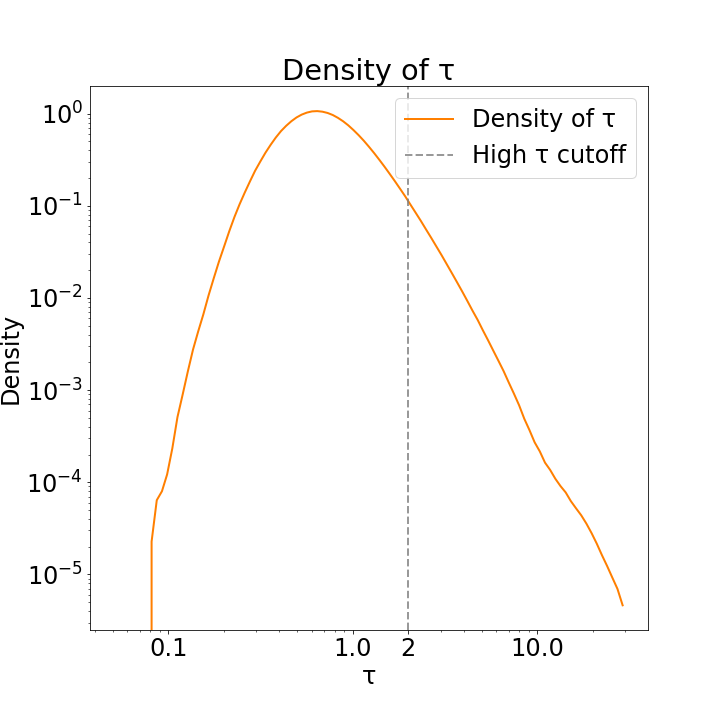}
    \caption{Density plot of $\tau$ values in pixels in our simulated
    \lya\ forest dataset. The vertical line at $\tau=2$ represents the 
    boundary between high and low $\tau$ values used in the analysis in Section \ref{results}.}
    \label{fig:tauhistogram}
\end{figure}

The spectra are generated with 4096 pixels each, but these
are rebinned into 512 pixels per sightline, in order
to approximate the resolution of spectra in the
Sloan Digital Sky Survey (SDSS, e.g., \citealt{lee2013}). The
pixel width is 90 $\kms$. In Figure \ref{fig:tauhistogram} we show the pdf of the
underlying $\tau$ values in the pixels. We can see that the mode of the distribution
is $\tau \sim 0.6$, which corresponds to $F\sim 0.5$. Nevertheless there are a significant number of pixels with high $\tau$.  Because these represent an interesting subset for our analysis (being close to saturated), we evaluate the
accuracy of the reconstruction separately for  high and low $\tau$ pixels.
The (arbitrary) boundary between the two sets of pixels we set to be at $\tau=2$
(which corresponds to $F=0.135$). This boundary is shown on Figure \ref{fig:tauhistogram}.

\begin{figure*}
    \includegraphics[width=0.7\textwidth]{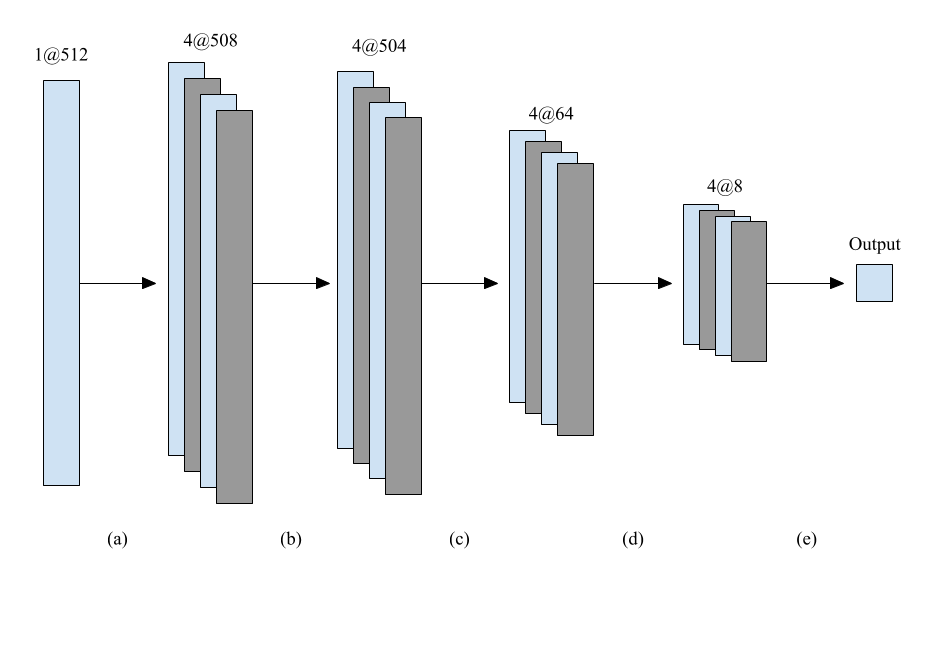}
    \caption{
      Architecture of the neural network used in this paper,
      which takes the flux as input and returns a prediction for
      optical depth $\tau$ at the center pixel. Layer (a) is a
      convolutional layer with kernel size of 5 and stride of 1. Layer (b) is a max
      pool layer which also has a kernel size of 5 and stride of 1. At the top of the figure, we write the number of channels and the size of feature map. For instance, 4@508 means the layer has 4 channels and 508 pixels. Layers (c), (d), and (e) are linear layers. Prior to
      layer (e), the array is flattened so that the final layer
      returns a single output. Layers (a), (c), and (d) go through the
      ReLU activation function, $f(x) = max(0,x)$. See Section
      \ref{NN} for more details.
    \label{fig:nnfig}}
\end{figure*}

We apply artificial noise to the $F$ values by adding to each pixel
a value randomly drawn from a Normal distribution with mean zero
and standard deviation $\sigma_{N}$. The signal to noise ratio is defined to be
\begin{equation}
S/N=\frac{\sigma_{F}}{\sigma_{N}},
\end{equation}
where $\sigma_{F}=0.635$ is the standard deviation of flux values averaged over
all spectra. We try three different S/N values during our analysis: 
10, 5, and 2.5. We explain further in Section \ref{preprocess}.

\begin{figure*}
  \begin{centering}
    \includegraphics[width=1.0\textwidth]{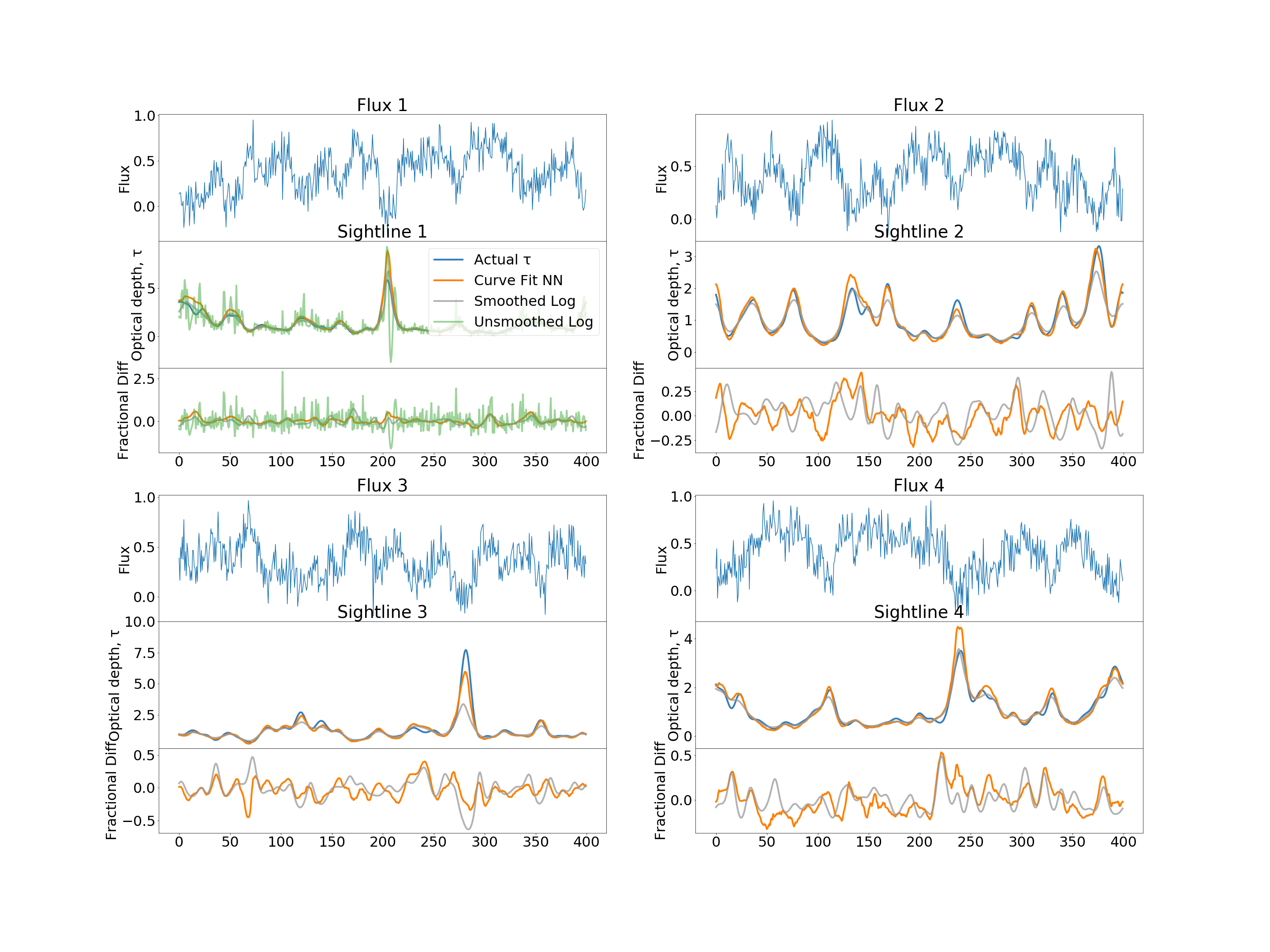}
    \caption{\lya optical depth $\tau$ predicted by the curve fit NN (detailed in Section \ref{scatterplots}) for four example sightlines. We also show
    results for alternative reconstruction methods involving taking the log and interpolating with  cubic splines
     (see Section \ref{comp}). In addition, we show the fractional difference between predicted $\tau$ and real $\tau$ for each method beneath the reconstructions. We first calculate flux, then add noise with a signal-to-noise ratio of 5 before using the four different reconstruction methods to predict $\tau$. The $x$-axis
    units are comoving $\hmpc$}
    \label{fig:four_sightlines}
\end{centering}
  \end{figure*}

\section{Method: Deep learning reconstruction of the Lyman-$\alpha$ forest}
\label{method}

\subsection{Data preprocessing}

\label{preprocess}
As explained in Section \ref{intro}, we  train a NN using our mock
datasets, and use it to recover the optical depth to \lya absorption, $\tau$,
from input values of flux $F$.

The data are split into training, test, and validation sets (the role of each will be explained in Section \ref{training} below). 60\% of the sightlines are assigned to the training set, and 20\% each to the validation and testing sets. The sightlines are arranged in a 2D array, where nearby sightlines are strongly correlated with each other. The sightlines that make up the validation and test sets come from opposite corners of the 2D array to ensure the test and validation sets are as independent as possible from the training data.

As mentioned above, we generate Gaussian noise at three different signal-to-noise ratios. While we generate new noise at every epoch for the training data, we create and save the noise for the validation and test sets for more consistent evaluation of the neural network and comparison to other reconstruction methods.

For each sightline, the data used for training are the mock observational datasets consisting of $F$ values, to which noise has been applied, and the optical depth $\tau$ values. The
latter represent the underlying physical quantities which the NN will learn how to derive from the former. Because we are primarily interested in the large scale structure of the forest, we smooth the $\tau$ values as part of our data preprocessing before passing them to the NN for training (the training flux values are unsmoothed). The smoothing is done with a Gaussian kernel of $\sigma$ equal to six pixels (4.7 $\hmpc$). 

\subsection{Neural network}

\label{NN}
Convolutional neural networks (CNNs) have revolutionized the field of computer vision in the last decade achieving  state-of-the-art performance on almost all computer vision tasks, such as image classification, object detection and semantic segmentation \citep{rawat2017deep}. CNNs provide the ability to solve non-linear inversion problems by learning spatial features that are invariant to translation and local distortions. Consequently they are being increasingly used in various fields other than computer vision including cosmology \citep{ribli2019weak,ravanbakhsh2017estimating}. Hence we choose to use a convolution neural network architecture to learn spatial features from flux input to predict optical depth, and the proposed architecture is displayed in Figure~\ref{fig:nnfig}. The inputs are the observed flux values, $F$, for the 512 pixels in the same sightline. The output is the prediction for the optical depth $\tau$ at the center pixel of a sightline (which we take to be the 256th out of the 512 pixels). To predict $\tau$ values for other pixels in the  sightline, we shift the pixels such that the position of output pixel is at the center of the sightline.  Each sightline can therefore be used for training 512 times, each time with a different pixel at the center. The periodic boundary conditions of the simulation are respected during this process.

In Figure~\ref{fig:nnfig}, we can see the different layers that the input is processed through. The first, (a) is a convolutional layer with kernel size of 5 pixels and stride of 1 pixel. This layer has 4 filters which it applies to the sightline, producing four-channel data from the originally one dimensional data. The next, layer (b) is a max pool layer: for discrete sets of five neighboring pixels, the layer outputs the maximum value using a stride of 1 pixel. This is done for each of the four channels individually. Layers (c), (d), and (e) are fully connected linear layers with decreasing numbers of outputs. Prior to layer (e), the array is flattened so that the final layer returns a single output. The outputs of layers (a), (c), and (d) pass through the Rectified Linear Unit (ReLU) activation function, $f(x) = max(0,x)$. The final architecture and hyperparameters are chosen by  experimentation using the validation dataset which is described in the following section \ref{sec:hyperparameters}.

As the NN is trained, the weights are adjusted based on minimization of a loss function. We use the mean squared error (MSE) as the loss function:
\begin{equation}
L(\hat{\tau}, \tau) = \frac{1}{n}\sum_n (\hat{\tau} - \tau)^2
\end{equation}

Here, $\tau$ is the actual $\tau$ value in a pixel and $\hat{\tau}$ is the neural network prediction for $\tau$, while $n$ is the batch size. The adjustment of weights is carried out using Adam optimizer \citep{kingma2014adam}, an efficient alternative to the standard stochastic gradient descent method. We also use L2 regularization, which adds a regularization term to the loss function. The goal is to decrease the network complexity and improve generalization. For L2 regulation, we use the decoupled weight decay regularization method \citep{loshchilov2017decoupled} that is part of the Adam optimizer implementation within the PYTORCH library \citep{NEURIPS2019_9015}. We set the weight decay parameter in L2 regularization to $5*10^{-4}$. The neural network is trained on 10,000 samples every epoch. The code base was written using the
PYTORCH library, and we make it publicly available  \footnote{\tt https://github.com/lhuangCMU/\linebreak deep-learning-intergalactic-medium. } to the research community.

\subsection{Hyperparameters}
\label{sec:hyperparameters}
We choose our proposed neural network architecture after training and evaluating various architectures across a range of hyperparameters on the validation set. We initially experiment with fully connected network architectures with varying numbers of hidden layers and units, but we find that it does not learn. Hence we switch to use convolutional neural network (CNN) architectures as motivated in the previous section. For the CNN architecture, we experiment with different kernel sizes between 3 to 5, output channels between 2 to 4, and linear layers between 2 to 5 to find the optimal architecture. For each of the architectures, we experiment with a range of learning rates between $10^{-3}$ and $10^{-6}$ for thorough comparison. Ultimately, we find that our proposed architecture described in Figure \ref{fig:nnfig} results in the most accurate predictions.

    Additionally, we experiment with different types of pooling layers including max, min and average pooling layers. Pooling is often performed after convolutional layers in order to reduce the spatial resolution of the feature maps and thus achieve spatial invariance to input distortions and translations. As we use a kernel size of 5, max pooling separates the data into groups of 5 consecutive pixels. For each of these groups, the maximum value within the consecutive pixels is taken as the output value for the group. Similarly, min pooling does the same using the minimum value, while average pooling takes the arithmetic average of the 5 pixel group. Some works \citep{scherer2010evaluation, jarrett2009best} have empirically shown that max pooling provides superior generalization and faster convergence leading to most state-of-the-art architectures using max pooling. However, another work \citep{boureau2010proceedings} focuses on theoretical analysis of max pooling and average pooling supplemented by empirical evaluations to conclude that the performance of either max or average pooling depends on the data and its features. Hence we conduct experimental analysis to evaluate the optimal pooling layer type for our dataset and architecture. We find that min and average pooling perform marginally better (within 3 standard deviations of Root Mean Squared Error, RMSE) for $\tau < 2$, but perform marginally worse (within 1 standard deviation) on higher values, $\tau > 2$, as compared to max pooling. Since the performance across different pooling layers is comparable for our dataset, we choose max pooling in our proposed architecture as it performs marginally better on the high $\tau$ values, which is the region of saturated flux data we are interested in. The details of RMSE and standard deviation calculation are provided in Section \ref{sec:stastitical_measures}.
    

We report the neural network's hyperparameters here for completeness. We use a learning rate of $10^{-4}$, batch size of 10,000, and train for 50,000 epochs. We use Adam optimizer, as implemented in the PYTORCH library, with betas equal to (0.9, 0.999), eps of $10^{-8}$, and weight\_decay of $5*10^{-4}$.

\subsection{Training}

\label{training}

In training, we randomly select 10,000 pixels from our training set, shifting their sightlines so that the selected pixel is at the center. We then add noise to the $F$ values. As explained in
Section \ref{mockobs}, we use a normal distribution for noise, with standard deviation determined by the desired signal-to-noise ratio.  The NNs are both trained and tested at a single signal-to-noise ratio. We do find, however that a NN trained using data with a S/N of 5 still outperforms comparison reconstruction methods at other signal to noise ratios (namely 2.5 and 10). We discuss further on this in Section \ref{discussion}.
The validation dataset is used to evaluate performance of the neural network and tune hyperparameters, while the test dataset is used for final results.

\subsection{Comparison methods for reconstruction}
\label{comp}

It will be useful to compare the NN reconstruction of $\tau$ from noisy $F$ values with other reconstruction methods. Looking at Equation \ref{Feq}, one can see that the simplest method would be a straight inversion, $\tau=-\ln{F}$. This is the first alternative reconstruction method that we try. 
Of course it is necessary in this case to deal with negative values of $F$. We do this using cubic spline interpolation. When a pixel $F$ value is negative, we initially ignore it, calculating the negative log of all positive $F$ values while saving their positions on the sightline. Once all positive $F$ values have a predicted $\tau$, we then use cubic spline interpolation with a periodic boundary to estimate $\tau$ for pixels with negative $F$.

\begin{figure*}
    \includegraphics[width=0.5\textwidth]{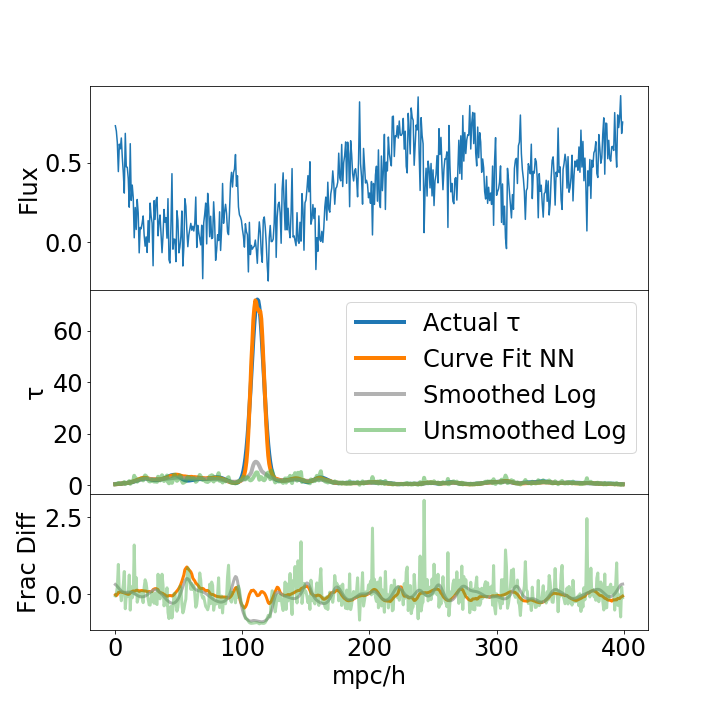}
    \caption{Reconstruction of  a sightline with high values of $\tau$. This sightline was selected for having the most pixels with $\tau$ greater than 50.  See the caption of Figure \ref{fig:four_sightlines} for more details.}
    \label{fig:hightausightline}
\end{figure*}

Our second comparison method (which gives better results) involves first smoothing the $F$ values with
a Gaussian kernel with a $\sigma$ of 6 pixels, and then computing  $\tau=-\ln{F}$. We label this
method Smoothed Input Log. In this case, there are fewer negative pixel values, but when there are, we again use cubic spline interpolation, as in the previous method.

\begin{figure*}
    \includegraphics[width=1.0\textwidth]{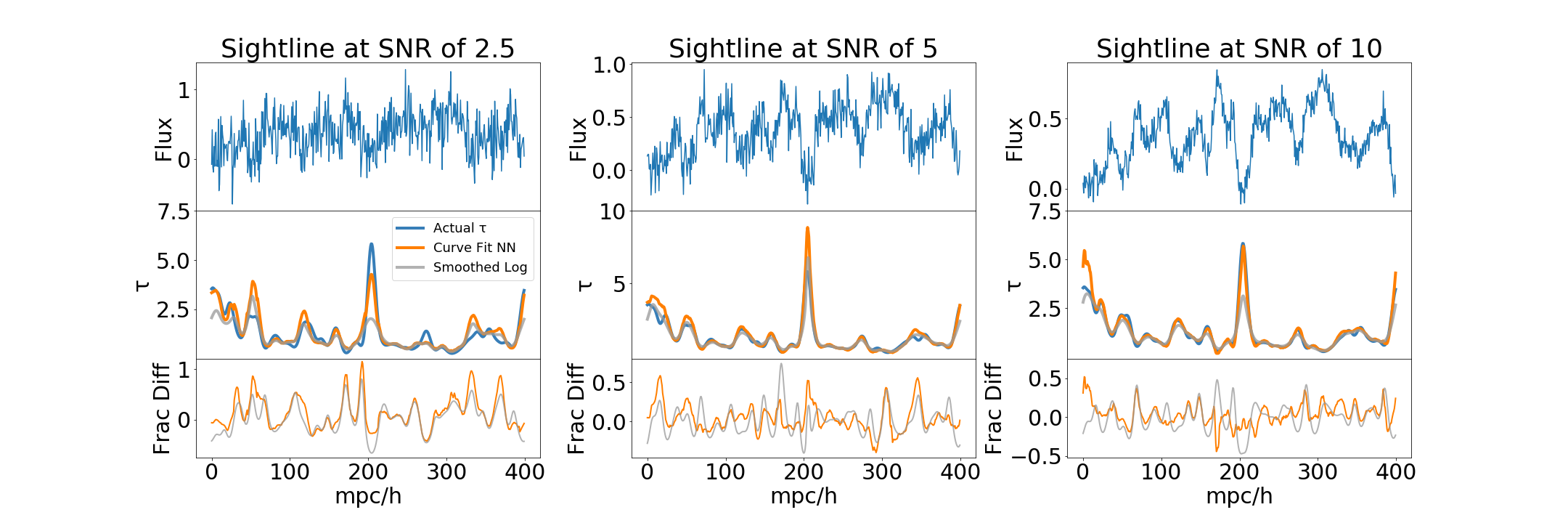}
    \caption{Reconstruction of the same sightline for three different noise levels. See the caption of Figure \ref{fig:four_sightlines} for more details.}
    \label{fig:threenoise}
\end{figure*}


\section{Results}

\label{results}
After training the NN for 50,000 epochs using the training data, and adjusting the hyperparameters using 
the validation dataset, we apply the NN reconstruction to the test dataset (which consists of 20\% 
of the sightlines). In this section we show some example sightlines as well as some statistical
evaluations of the fidelity of the reconstructions.

\begin{figure*}
    \begin{subfigure}[b]{\columnwidth}
        \includegraphics[width=1.0\textwidth]{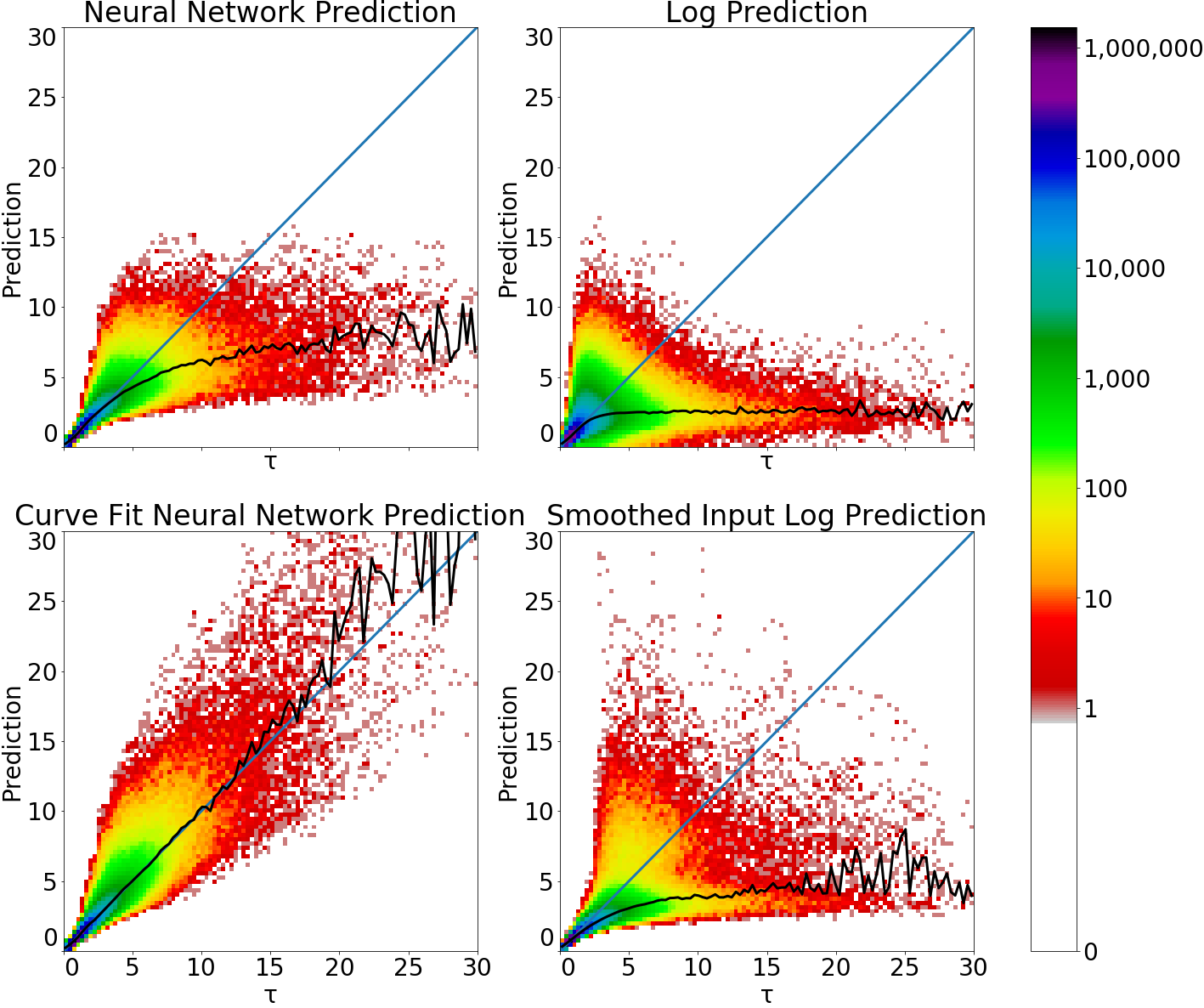}
    \end{subfigure}
    \begin{subfigure}[b]{\columnwidth}
        \includegraphics[width=1.0\textwidth]{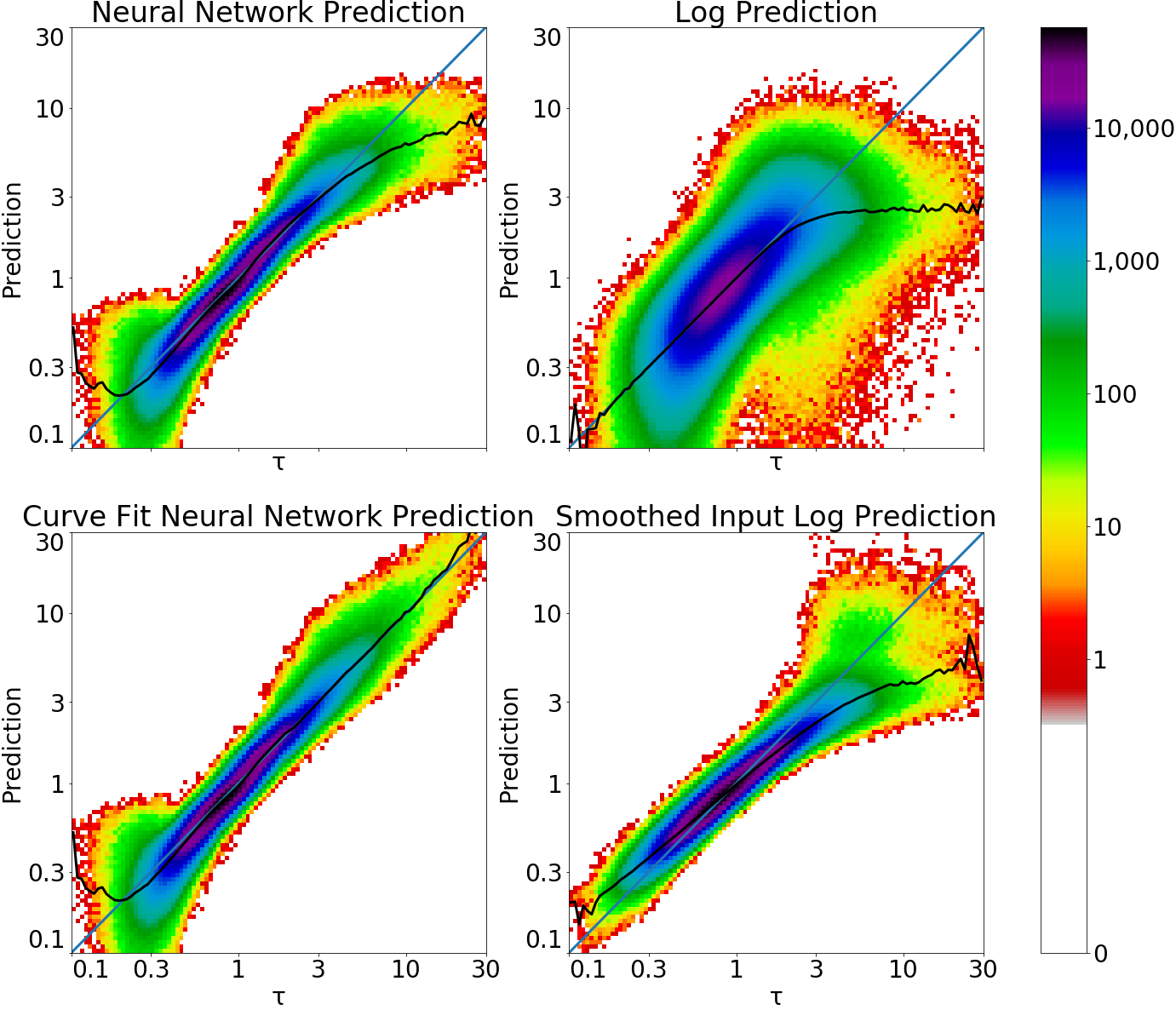}
    \end{subfigure}
    \caption{Scatter plots of the reconstruction predictions for the 
    \lya optical depth $\tau$ as a function of the true $\tau$ in each pixel. Each panel shows a different reconstruction technique. Clockwise
    from top left these are, (a) NN prediction; (b) Log inversion of the unsmoothed $F$ data, including cubic spline
    interpolation; (c) Log inversion of the smoothed $F$ data, including cubic spline interpolation; and (d) NN prediction adjusted with curve fit to high $\tau$ pixels.
    The line in black shows the median prediction for each bin of the real tau value. The right subfigure shows the same scatter plots as the left, but with log scaling in the axes.
    }
    \label{fig:scatter}
\end{figure*}

\subsection{Example sightlines}
In Figure \ref{fig:four_sightlines} we show results for four randomly chosen sightlines. We show the input noisy $F$ values as a function of distance along the sightline
in the top panel in each case. All of the examples in this plot are for an 
input spectra with S/N = 5. Underneath the $F$ panel in each case, we show the actual
$\tau$ values, as well as the results of the curve fit NN reconstruction (detailed in Section \ref{scatterplots}), and the Smoothed Input Log inversion. 
For only one of the panels (the top left) we show the results of the direct log inversion ("Unsmoothed Log", in green), but do
not show it in the others because it obscures the other results. We also show the fractional difference between predicted $\tau$ and real $\tau$ for each method beneath the reconstructions. The fractional difference is calculated as $(\tau_{pred} - \tau_{real})/\tau_{real}$.

We can see that the NN has learned to reconstruct the $\tau$ curve from the noisy flux quite well. The general nature of the fluctuations is reproduced, even in regions where the $F$ values become
significantly negative due to noise. The Smoothed Input Log reconstruction also works reasonably well, although appears to underpredict in the high $\tau$ regions.

In Figure~\ref{fig:hightausightline}, we show the situation for a sightline with very high $\tau$ values, again with S/N = 5. This sightline was chosen because it had the most pixels with $\tau>50$. We
can see that there is a significant region, about $100 \hmpc$ in width, where the flux values are
roughly consistent with zero given the noise. The direct inversion method shown in green does not capture
any of the high $\tau$ structure in this region. The Smoothed Log method does find a bump with $\tau \sim 10$ in the right place, but the NN is able to use the information on $F$ surrounding the high $\tau$ pixels to reconstruct a reasonable likeness of the "hidden" $\tau$ in the most absorbed region.

The previous plots showed results for a moderate noise level, S/N=5. In 
 Figure~\ref{fig:threenoise}  we show the flux and the $\tau$ predictions across the different levels of noise we have tried, where the S/N values are equal to 2.5, 5, and 10. In this case, we show the same sightline in each case, only the noise
 being different. We do not show the Unsmoothed Log reconstruction to avoid obscuring the other lines.
We can see that there are significant differences in the small scale structure of $\tau$  reconstructed by both the NN and the Smoothed Log in the low S/N case, although both recover the
largest peak quite well. As the S/N increases, the fidelity becomes markedly better, with the largest qualitative improvement between S/N=2.5 and S/N=5. The statistical evaluation of the accuracy of this method for different S/N is carried out in the Section~\ref{sec:stastitical_measures}.

\subsection{Scatter plots}
\label{scatterplots}

We now move to comparisons of results from all pixels. We show scatter plots of the $\tau$ predictions from the NN and other
reconstruction methods vs the true $\tau$ values in Figure 
\ref{fig:scatter}.

The NN prediction is in the top left panel, and we can see that the $y=x$ line appears to pass through the center of the point cloud for $\tau$ values below 10. 
We see however that the NN prediction appears to be non-linear for $\tau$ values that are higher than this. For example, the $\tau$ prediction never rises above $\tau=16$, although there are pixels in the spectra which have higher $\tau$ values.

The neural network likely has difficulty predicting high $\tau$ for two reasons. Pixels with high $\tau$ values will have a low flux, so any added noise can have a large impact. The second reason is due to a class imbalance problem \citep{buda2018systematic}, as pixels of high $\tau$ are rarer than pixels of low $\tau$, with only 0.01\% of pixels being above $\tau = 20$. We attempt to deal with the class imbalance problem by adjusting the loss function to evaluate loss differently for higher values of $\tau$. We divide pixels into 3 bins; $\tau < 2$, $2 \leq \tau < 30$, and $\tau \geq 30$. We calculate the proportion of pixels in each bin then take the weighted average of square errors: \begin{equation} L(\hat{\tau}, \tau) = \frac{1}{n}\bigg(\sum_{\tau < 2} \frac{(\hat{\tau} - \tau)^2}{b_1} + \sum_{2 \leq \tau < 30} \frac{(\hat{\tau} - \tau)^2}{b_2} +  \sum_{\tau \geq 30} \frac{(\hat{\tau} - \tau)^2}{b_3}\bigg)\end{equation}
Where $b_i$ refers to the proportion of pixels in the entire training set that are in each bin and n is the batch size, which we chose to be 10,000. The numerical values for the bin proportions are as follows; $b_1 = 0.906, b_2 = 0.094, b_3 = 2.12*10^{-5}$.

With this method, we increase the loss in bins of high $\tau$ according to the proportion of pixels in each bin. This method was unsuccessful in increasing accuracy, with a RMSE value of 0.533 at a signal-to-noise ratio of 5, which was 25\% higher than using mean square error as the loss function. We also attempted using two neural networks to predict one sightline, with one network predicting low values of $\tau$ and the other predicting high values of $\tau$. These models' architectures are the same architecture we use in this paper. This method was also unsuccessful in increasing accuracy.

Another possibility is to directly address the class imbalance problem by training the network on sightlines from a range of redshifts so that the network has more training data with high $\tau$ values. We leave exploring this method to future work.

Even though the number of extremely high $\tau$ pixels is small
(only 1.16\% of pixels have true $\tau>10$), in order to achieve higher accuracy, we use a curve fitting algorithm on the ratio of actual $\tau$ to predicted $\tau$ where the actual $\tau$ is greater than or equal to 2. For these datapoints, we use the actual to prediction ratio as our y-values and the actual $\tau$ as our x-values. By fitting a curve to these points, we construct a function of actual $\tau$ that outputs the ratio between actual $\tau$ and predicted $\tau$. In order to correct our neural network predictions, we multiply each point in the scatterplot by the ratio given by the function and the actual $\tau$ value of the pixel. 

When fitting a cubic function to these ratios, we find parameters that minimize the residuals between the cubic function and these datapoints using the Levenberg-Marquadt optimization algorithm. The resulting cubic function is $r=-0.000078\cdot x^{3}+0.0046\cdot x^{2}+0.047\cdot x+0.81$, where r is the ratio between actual $\tau$ and predicted $\tau$. Because the neural network's prediction is linear for low $\tau$, we don't modify those points. In future work, we will investigate whether the NN can be trained to do better on the highest $\tau$ points, but do not do this here, in order to keep the NN part of our algorithm simple.

The result of including a curve fit to the predictions is shown in the  "Curve Fit Neural Network Prediction" panel in Figures~\ref{fig:scatter}. We do not apply the same method to the analytical method Smoothed Input Log because its prediction at the highest $\tau$ values is approximately symmetric about $y=x$, and we find that curve fitting would not increase accuracy significantly. 

The results from the Smoothed Input Log reconstruction are in the bottom right panel of Figure \ref{fig:scatter}. We can see both that the scatter extends significantly wider than for the NN method, and that there is curvature in the mean relation even for  values as low as $\tau \sim3$. As mentioned above, for higher $\tau$ values (above $\tau=15$) there is not evidence for curvature but the scatter is extremely high. The Unsmoothed Log Prediction (top right panel) is not biased at low $\tau$, but has
visually much worse scatter.

\subsection{Statistical measures}
\label{sec:stastitical_measures}

We have seen that the neural network appears to qualitatively outperform our alternative reconstruction methods,
and have seen some examples of sightlines with different levels of signal to noise. We now evaluate the
performance quantitatively, by comparing the reconstructed $\tau$ values in sightlines to the true $\tau$ values. Again, the results are from predictions on the test set, which the neural network has not trained on.
One measure of the accuracy of the reconstruction is the Root Mean Squared Error, RMSE, defined as
\begin{equation}
\label{rmse}
{\rm RMSE}= \sqrt{\frac{1}{n}\sum_{i=1}^{n} (\hat{\tau_{i}} - \tau_{i})^2},
\end{equation}
where the sum is over the $n$ pixels in the test dataset,  $\hat{\tau_{i}}$ is the reconstructed optical depth
in  pixel $i$ and $\tau_{i}$ is the true value.

Our second measure of the accuracy is the fractional error, which we define to be RMSE$/{\overline{\tau}}$, where ${\overline{\tau}}$
is the mean optical depth for the particular dataset being evaluated. The different datasets are either the full range of pixels in spectra, or the high $\tau$ pixels (with $\tau > 2$),
or those with low $\tau$ ($\tau < 2$)
Across the entire data set, mean ${\overline{\tau}}$ is 1.107. For $\tau < 2$, mean ${\overline{\tau}}=0.543$, and for $\tau \geq 2$, ${\overline{\tau}}= 5.468$.

There are three levels of stochasticity to the RMSE values. The first comes from the noise, the second comes from the initial weights of the neural network, and the third comes from the source of our sightlines. In order to capture two of the three levels of stochasticity, the RMSE values in Tables~\ref{tab:RMSE10},~\ref{tab:RMSE5}, and~\ref{tab:RMSE2.5} are the averages of 7 neural networks with different initial weights predicting $\tau$ with different generated noise. The comparison reconstruction RMSE values are also an average over 7 different sets of randomly generated noise.
The standard deviations are calculated from the seven different reconstructions in each category using the following formula: \begin{equation} \sigma = \frac{\sqrt{\sum (x-\bar{x})^2}}{n-1} \end{equation} Here, $x$ is the RMSE value, $\bar{x}$ is the average RMSE value for the reconstruction method at a given signal-to-noise ratio and $\tau$ range, and n is the number of samples, which is 7 here.

\begin{table}
    \centering
    \caption{RMSE of our neural network's prediction vs log prediction over our test set, with a signal-to-noise ratio of 10. The RMSE is split into three sections, where we calculate RMSE for the total test dataset, just for high values of $\tau$, and just for low values. We define $\tau \geq 2$ to be a high value of $\tau$.}
    \label{tab:RMSE10}
    \begin{tabular}{llll}
        \hline
        Name & RMSE$_{\text{total}}$ & RMSE$_{\text{high}}$ & RMSE$_{\text{low}}$ \\
        \hline
        Curve Fit NN & 0.285 $\pm$ 0.01 & 0.882 $\pm$ 0.05 & 0.091 $\pm$ 0.02 \\
        Neural Network & 0.330 $\pm$ 0.01 & 1.036 $\pm$ 0.03 & 0.091 $\pm$ 0.02 \\
        Log & 0.620 $\pm$ $7\mathrm{e}{-4}$ & 1.908 $\pm$ $2\mathrm{e}{-3}$ & 0.214 $\pm$ $2\mathrm{e}{-4}$ \\
        Smooth Input Log & 0.511 $\pm$ $2\mathrm{e}{-3}$ & 1.620 $\pm$ $8\mathrm{e}{-3}$ & 0.124 $\pm$ $5\mathrm{e}{-5}$ \\
        \hline
    \end{tabular}
\end{table}

\begin{table}
    \centering
    \caption{RMSE values for SNR of 5 (see Table \ref{tab:RMSE10} caption for details)}
    \label{tab:RMSE5}
    \begin{tabular}{llll}
        \hline
        Name & RMSE$_{\text{total}}$ & RMSE$_{\text{high}}$ & RMSE$_{\text{low}}$ \\
        \hline
        Curve Fit NN & 0.342 $\pm$ $4\mathrm{e}{-3}$ & 1.012 $\pm$ 0.01 & 0.151 $\pm$ $8\mathrm{e}{-3}$ \\
        Neural Network & 0.423 $\pm$ $3\mathrm{e}{-3}$ & 1.296 $\pm$ 0.01 & 0.151 $\pm$ $8\mathrm{e}{-3}$ \\
        Log & 0.800 $\pm$ $1\mathrm{e}{-3}$ & 2.192 $\pm$ $4\mathrm{e}{-3}$ & 0.455 $\pm$ $3\mathrm{e}{-4}$\\
        Smooth Input Log & 0.538 $\pm$ $4\mathrm{e}{-3}$ & 1.696 $\pm$ 0.01 & 0.141 $\pm$ $4\mathrm{e}{-4}$\\
        \hline
    \end{tabular}
\end{table}

\begin{table}
    \centering
    \caption{RMSE values for SNR of 2.5 (see Table \ref{tab:RMSE10} caption for details)}
    \label{tab:RMSE2.5}
    \begin{tabular}{llll}
        \hline
        Name & RMSE$_{\text{total}}$ & RMSE$_{\text{high}}$ & RMSE$_{\text{low}}$ \\
        \hline
        Curve Fit NN & 0.430 $\pm$ $7\mathrm{e}{-3}$ & 1.044 $\pm$ 0.08 & 0.299 $\pm$ 0.03 \\
        Neural Network & 0.560 $\pm$ $8\mathrm{e}{-3}$ & 1.570 $\pm$ 0.03 & 0.299 $\pm$ 0.03\\
        Log & 1.076 $\pm$ $8\mathrm{e}{-4}$ & 2.529 $\pm$ $3\mathrm{e}{-3}$ & 0.782 $\pm$ $4\mathrm{e}{-4}$ \\
        Smooth Input Log & 0.674 $\pm$ $7\mathrm{e}{-3}$ & 2.100 $\pm$ 0.02 & 0.208 $\pm$ $1\mathrm{e}{-3}$ \\
        \hline
    \end{tabular}
\end{table}

\begin{figure*}
    
    \begin{subfigure}[b]{\columnwidth}
        \includegraphics[width=\columnwidth]{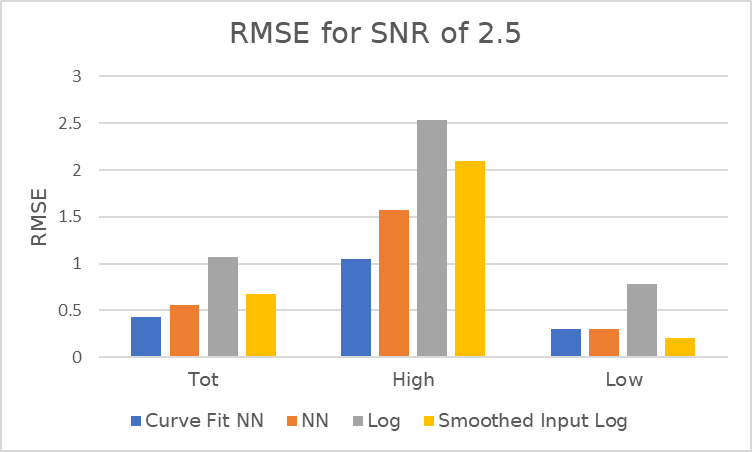}
        \includegraphics[width=\columnwidth]{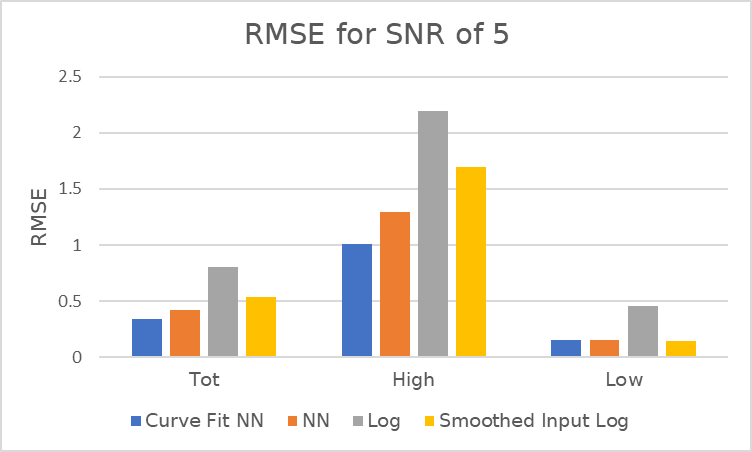}
        \includegraphics[width=\columnwidth]{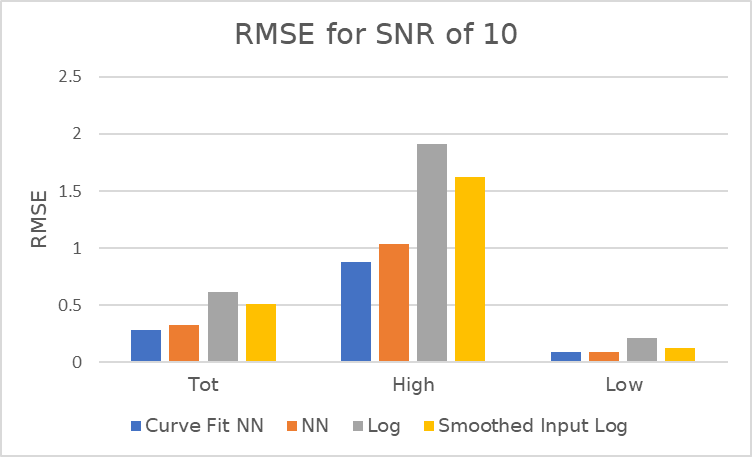}
        \caption{RMSE}
        \label{fig:noiseRMSE}
     \end{subfigure}
    \centering 
     \hfill %
     \begin{subfigure}[b]{\columnwidth}
        \includegraphics[width=\columnwidth]{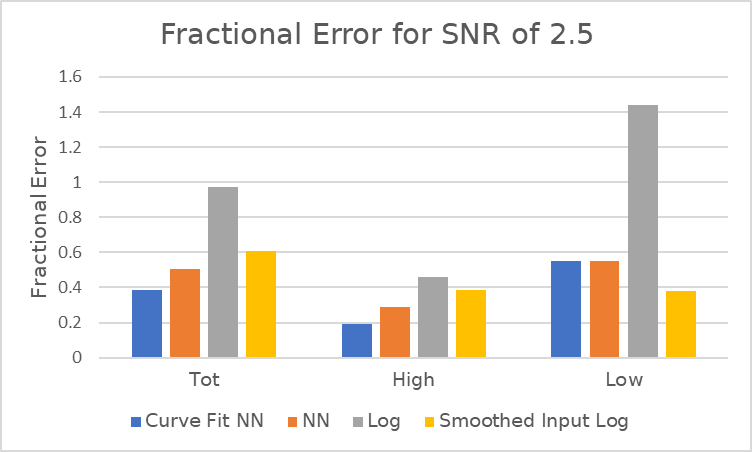}
        \includegraphics[width=\columnwidth]{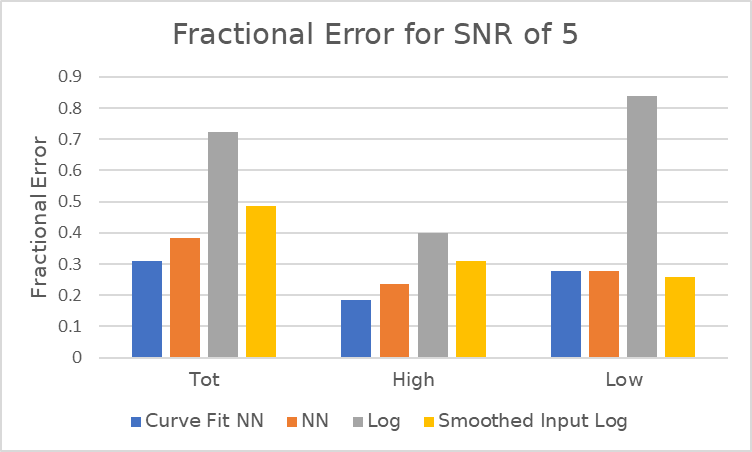}
        \includegraphics[width=\columnwidth]{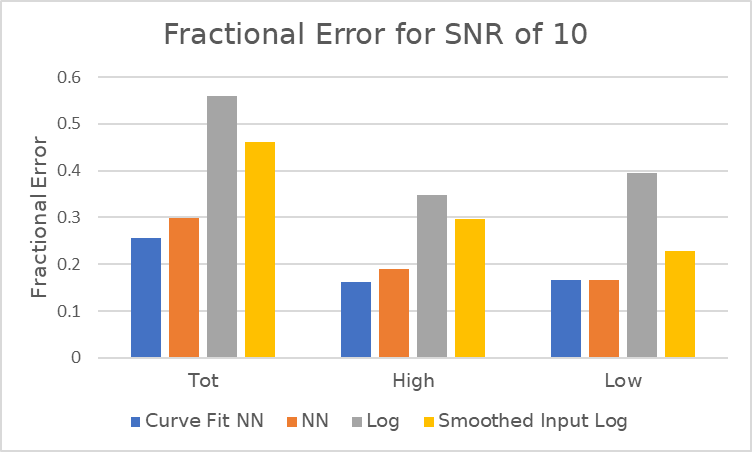}
        \caption{Fractional Error}
        \label{fig:noiseFrac}
     \end{subfigure}

    \caption{RMSE and Fractional Error for different methods across different noise levels, separated further into three groups; all $\tau$ values; only high $\tau$ values; and only low $\tau$ values. The four methods are neural network prediction, curve fit neural network prediction (where the output is fit onto $y = x$), log prediction, and smoothed input log prediction (where the input flux is smoothed with a Gaussian kernel size of 6 pixels). The RMSE is calculated using Equation \ref{rmse}. The fractional error 
    is defined to be RMSE divided by mean $\tau$ in each group.}
    \label{fig:noiseResults}
\end{figure*}

We present our results in Tables \ref{tab:RMSE10}-\ref{tab:RMSE2.5} and in
graphical form in Figures \ref{fig:noiseRMSE} and \ref{fig:noiseFrac}. A
quick glance reveals that in these figures that the blue bar, the NN
adjusted by curve fitting has the lowest RMSE and Fractional error
in all cases, except for the $\tau < 2$ results for S/N of 2.5 and 5. The
improvement over the raw log transformation is significant for the
total of all pixels, and for $\tau < 2$, varying from a
factor of 2.1 to 2.5, with no variation for different S/N.
If we compare instead to the smoothed input log reconstruction, we find
that that the curve fit neural network improves the reconstruction by a factor
of between 1.6 and 2.0 for all and high $\tau$ pixels.

The curve fitting addition to the NN
makes the most difference for
low S/N and high $\tau$ pixels. There is no
difference for $\tau <2$. The improvement over the NN on  its
own varies from a factor of 1.1 to 1.5. Apart from the $\tau<2$ low S/N
results mentioned above, the NN without curve fitting is significantly
better than the smoothed input log reconstruction.
When considering the accuracy on a pixel by pixel basis, the fractional
error (Figure \ref{fig:noiseFrac}) is useful. We can see that we can
aspire to a fractional error on $\tau$ reconstruction for $\tau >2 $
using the curve fit NN
of less than 20\%, for all S/N levels tested.

\section{Summary and discussion}
\label{summary}

\subsection{Summary}

We have set up a neural network to train on 1D \lya forest datasets from
simulations.The aim is to use the trained NN to recover
underlying \lya optical depth values from noisy and saturated
transmitted flux data in quasar spectra. The NN has an architecture which
includes both convolutional and fully connected layers. We have trained
the network using spectra from hydrodynamic simulations of a CDM
cosmology. The NN has been applied to a test dataset, and its accuracy
evaluated statistically using the root mean square difference between the
reconstructed and true $\tau$ values in the simulation. We have compared
the NN reconstruction to straightforward logarithmic inversion of the noisy
flux data (including spline interpolation of $\tau$ through saturated
regions) and also logarithmic inversion of the smoothed flux data.
Our findings are as follows:
\begin{itemize}
  \item The curve fit neural network is at least twice as accurate as the naive log reconstruction method.
  \item Curve fitting decreased the neural network's RMSE by 15\% on average.
  \item The curve fit neural network outperforms all the other methods except the Smoothed Input Log method for low values of $\tau$ ($\tau$ < 2) where the signal-to-noise ratio is 2.5 or 5.
\end{itemize}

\subsection{Discussion}

\label{discussion}

Although we have concentrated on the simplest task in this paper, inversion of Equation \ref{Feq} for
noisy and saturated data, it should be relatively straightforward in principle to apply the 
same techniques to reconstruct other quantities from the transmitted \lya\ forest flux, $F$. The 
simulations include information on the underlying physical quantities relevant to $F$, such
as the baryonic and dark matter density, temperature and velocity fields. We leave testing
such reconstructions to future work, but we note that some quantities such as the velocity field
may be difficult to infer from individual 1D sightlines, as they are generated by the matter 
distribution in three dimensional space. It will nevertheless be interesting to see how much
can be recovered from one dimension only. All reconstructed quantities will of course be dependent
on the simulations and model used for training the NN (we return to this below). For example, little
direct information on the gas temperature is available from the low resolution spectra we have considered
so far (thermal broadening occurs on too small a scale), but a NN would presumably recover a physically
reasonable but very model dependent temperature indirectly from the relationship between
temperature and density in the IGM (\citealt{hui97}). Recent work on a similar theme, but in 
three dimensions is that of \cite{hong20} who have used hydrodynamic
simulations of galaxy formation to train a NN to reconstruct the dark matter distribution from
galaxy positions and velocities.

Having only trained our NN on one simulation, the answers that it returns are
likely to be strongly dependent on that training set. We have carried out tests using mock
data with different S/N ratios (training with a different S/N than the test data), and
find reasonable results, but it would be very interesting in future work to try
training the NN with data from different redshifts or cosmologies from the test data. 

Another issue related to the finite size of the training dataset is that there will be
rare events which could be underrepresented, such as large fluctuations in the
optical depth. There will also be features in real data which are not included in the
simulations, depending on their level of sophistication. For example we have not
included damped \lya\ lines in our mock datasets, or metal lines. In principle these
could be added to training data, as simulations exist which make predictions
for them (e.g., \citealt{pontzen08}). The physics involved (including galaxy and star formation)
is however more uncertain and less likely to resolved
in the simulations than the physics leading to the majority of the \lya optical depth.

We have compared the NN reconstruction method with two other methods for inferring
the optical depth from the flux. It is of course possible that other methods could be imagined which
have better performance. For example, in one method, we smooth the flux before log inversion and spline
interpolation. One
could imagine using some more sophisticated denoising such as L1 trend filtering \citep{politsch20_1,politsch20_2}
before log inversion. Physical reconstruction modeling could also be tried, which uses
the physics of the intergalactic medium in simulations to go from flux
to physical quantities. Examples include \cite{nusser99}, and \cite{muller20}. Other machine learning techniques have been applied to similar 
problems in absorption line data, for example use of a genetic algorithm to model
data with multiple metal line species (\citealt{lee20}),
or the use of conditional neural spline flows to predict the quasar continuum on the red side of the
quasar \lya\ line from blue side data \citep{reiman20}.

Our particular NN approach works better than the alternatives we have tested, except for the highest optical depth regions $\tau \simgt 15$, where the scatter is low
but there is a bias. These correspond
to an extremely small percentage of pixels, but nevertheless it would be very useful
to improve the NN there. We have investigated changes in NN hyperparameters, but
have not been able to simply improve the NN performance in these regions. 
We have
instead adopted a curve fit approach to the highest $\tau$ pixels, which, like
the NN uses information from the simulations. The combined NN and
curve fit approach does yield good results at high $\tau$, making use
of the fact that the NN is able to reduce the scatter even though its results are
biased. We leave a comprehensive effort to improve the NN in these regions to 
future work.

Another open question is how the NN is making its predictions for $\tau$. The flux in an
entire simulated spectrum (spanning 512 pixels and 400 $\hmpc$) is used by the NN as an input. In
future work, we plan to investigate the response of the NN and how it is using the input
information, for example weighted by pixel distance. The log inversion techniques use only single 
pixel information in unsaturated regions, but signal over longer distances in the spline
interpolation part of the algorithms. It will be interesting to compare the dependence of
the NN algorithm on distance of the farthest data used from the predicted pixel.

 We have approached this paper from the point of view that solving the inverse problem (in this example of observed flux
 to underlying optical depth) is an interesting intellectual
 exercise. One should obviously also ask however how useful our DL solution actually is, what its limitations might be.
Different use cases can be imagined, but they will likely all be dependent on the model used in training, unless significant
testing (for example with different simulations) shows how more general conclusions can be inferred. We indulge in a limited
amount of speculation here.
If we are testing a particular model (for example $\Lambda$CDM with specific parameters), we
could use simulations of that model for training and then compare statistics of the reconstructed fields (e.g., 
temperature, density) to see if they are
consistent with the original model. This would allow testing  using statistical
measurements of quantities which are not directly observable. 
In the case of \lya optical depth $\tau$ studied in this paper, we could imagine measuring the clustering of $\tau$,
including perhaps higher order statistics. Whether these would actually have more discriminatory power than statistics
of measurable quantities such as the flux $F$ is debatable, but at the very least they may offer different ways of weighting the data (see \citealt{mccullagh16} and related works for other approaches). For
example the S/N of \lya BAO measurements may improve (or not) if the observations are transformed to a $\tau$ field or a density field first. 

Certainly, in the case of the \lya forest there is increasing interest in the use of interpolation techniques to 
construct three dimensional maps from arrays of one dimensional spectra \citep{pichon01,horowitz19,newman20}). Instead of producing a 3D flux
field, one could use the NN reconstruction to make 3D $\tau$, temperature, or density fields. One use of reconstructed sightlines or maps could be to use in cross-correlation with other data. The \lya\ forest has a low bias factor (the
ratio of $F$ fluctuations to matter fluctuations), with $|b|\sim0.2$ \citep{slosar11}, and transforming to a variable with a
higher $|b|$ such as $\tau$ could increase the S/N of \lya\ forest - \lya\ emission cross-correlations
(e.g., \citealt{croft18}), for example.  Because the DL reconstruction appears to work significantly better on noisy data
than smoothing does, one could imagine using it to remove noise artifacts, or perhaps even set the unobserved quasar continuum
level (by training on mock data with varied continua).  

We have seen that NN are able to learn the relationships between complex physical quantities in simulations. In the case
of the \lya forest this can be used to carry out model dependent reconstruction from observables. As with many applications
of Artificial Intelligence techniques, the uses and limitations are not all yet apparent, but it is obvious that 
there is much of promise that should be studied further.

\subsection*{Acknowledgments}
RACC is supported by  NASA ATP 80NSSC18K101,  NASA ATP NNX17AK56G, NSF NSF AST-1909193, and
a Lyle Fellowship from the University of Melbourne. This work was also supported by the NSF AI Institute: Physics of the Future, NSF PHY- 2020295. We would like to thank the referee for their comments and suggestions for future work.

\subsection{Data availability}
The data underlying this article are accessible through the code repository.\footnote{\tt https://github.com/lhuangCMU/\linebreak deep-learning-intergalactic-medium. }.

\bibliographystyle{mnras}
\bibliography{ref} 

\end{document}